\documentclass{article}
\usepackage[affil-it]{authblk}
\usepackage[english]{babel}

\usepackage[letterpaper,top=2cm,bottom=2cm,left=3cm,right=3cm,marginparwidth=1.75cm]{geometry}

\usepackage{amsmath}
\usepackage{graphicx}
\usepackage[colorlinks=true, allcolors=blue]{hyperref}

\title{ Interplanetary Laser Tri-lateration Network: simulation with INPOP planetary ephemerides}
\author{A. Fienga%
  \thanks{Electronic address: \texttt{agnes.fienga@oca.eu}; Corresponding author}}
\affil{Universit\'e C\^ote d'Azur, Observatoire de la C\^ote d'Azur, CNRS, 250 av. A. Einstein, 06250 Valbonne, France \\ IMCCE, Observatoire de Paris, PSL University, CNRS, 77 av. Denfert-Rochereau, 75014 Paris, France}

\begin{document}
\maketitle

\begin{abstract}
This study is done in the context of the project titled  Interplanetary Laser Tri-lateration Network (ILTN) proposed by \cite{2018P&SS..153..127S} and investigated more in details by \cite{2022P&SS..21405415B} and \cite{2022P&SS..21505423B}. The original idea was to propose interplanetary measurements (in this case between Venus, Mars and the earth) as a way to measure the solar system expansion.  But some recent interests on the measurement of asteroid masses and more generally the study of the mass distribution in the outer solar system appear with the ILTN.
In this work, we are investigating how different possible configurations of interplanetary measurements of distances can be introduced in planetary ephemeris construction and how they improve our knowledge of planet orbits and other related parameters.

\end{abstract}

\section{Introduction}

We presents here results of simulations based on the planetary ephemerides INPOP21a \cite{2021NSTIM.110.....F} including Venus-Earth, Mars-Earth and Venus-Earth tracking ranges with accuracies from 0.1~mm to 10$^{-3}$~mm. Are considered, the full 401 INPOP21a parameters as well as  the  $\frac{\dot{\mu}}{\mu}$ where $\mu$ is the product of the gravitational constant G and the mass of the Sun. The results show that the most critical input come from the Mars-Venus observations improving by 50$\%$ the accuracy of  $\frac{\dot{\mu}}{\mu}$ obtained with only Venus-Earth, Mars-Earth  measurements for the same instrumental configuration.

\section{Simulation set}

We use INPOP21a \cite{2021NSTIM.110.....F} as reference planetary ephemerides (PE). We consider three durations of mission : 5 years, 2.5 years and 1 year and for each duration, three configurations: only range between Venus and the Earth and Mars  and the Earth with an un-precedent accuracy of 0.1~mm (case 1), full triangulation between Venus, Earth and Mars at  0.1~mm adding the Venus-Mars distance to case 1 (case 2), and the full triangulation with  10$^{-3}$~mm accuracy (case 3). For each case, we suppose 1 normal point per day, the given accuracies described below being for this normal point. We also consider a very conservative limit of 30 degrees for the minimal angular distance (or separation angle) between the planets and the Sun (SEP angle). 

\section{Configuration A: 5 years}

In order to be conservative, we consider the same filter applied to the present radio tracking observation for the planetary ephemeris (PE) construction: observations are rejected with the SEP is smaller than 30 degrees.
By considering the PE limits in term of SEP angles, we reduce the number of observations by  21$\%$. However, the results are very similar.
The results obtained with the SEP filter are plotted on Fig. \ref{casAa} in terms of ratio of the covariances for each of the PE parameters obtained in including the Case 1, 2 and 3 observations to the INPOP21a covariance. Comparing Case 1 (dots) and 2 (cross) from Fig. \ref{casAa}, it appears that the introduction of the Venus-Mars distances brings an improvement, in average, of about 1.2 for the inner planets and 1.3 for the asteroid masses. 
The semi-major axis of the planet orbits are usually the most sensitive to the addition of the Venus-Mars observations, which was expected. We also see that globally the asteroid masses are more constraints with the addition of the Venus-Mars measurements.
For the outer planets and the mass of the TNO ring,  the impact is more negligible but it is of about a factor 2 the gravitational mass of the Sun and its oblateness. 
With the Case 3 (10$^{-3}$~mm accuracy), the improvement is even more important with a gain of 2 orders of magnitude for all the asteroid masses, Venus and EMB orbits, the Earth Moon mass ratio, the mass of the Sun and its oblateness. For Mars, Jupiter, Saturn and the TNO ring mass, the improvement is more limited, to 1 order of magnitude. For Mercury and the outer planets, the improvement is less than 1 order of magnitude.
On Table \ref{tab:log}, are also considered here the impact of Cases 1,2 and 3 on the covariance of the estimation of $\frac{\dot{\mu}}{\mu}$. The introduction of the Venus-Mars distances does not impact strongly the measure of $\frac{\dot{\mu}}{\mu}$ (bu a factor 1.8) but the use of 10$^{-3}$~mm accuracy leads to an improvement of about 3 orders of magnitude (Table \ref{tab:log} in comparison with  the present accuracy and by a factor 50 relative to the measurements at 0.1~mm accuracy. 

\begin{figure}
\includegraphics[scale=0.5]{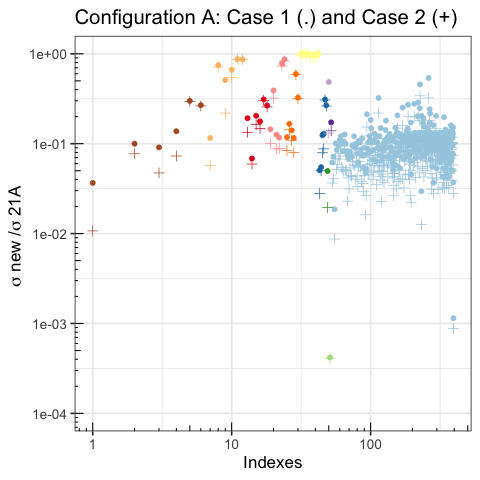}\includegraphics[scale=0.5]{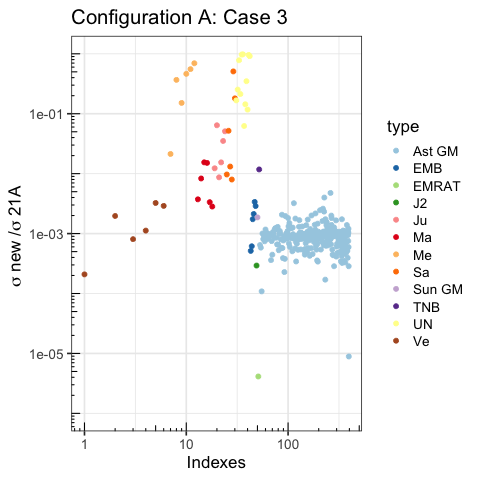}
\caption{Ratio of the uncertainties obtained for the 401 INPOP21a parameters  ($\sigma_{21A}$) and with the addition of ILTN observations for the different cases over 5 years with 1 normal point per day considering the separation angle between the sun and the planets: Case 1 is for Venus-Earth (VE) and Mars-Earth (ME) with an accuracy of 0.1~mm, Case 2 is for Venus-Earth, Mars-Earth and Venus-Mars (VM) with an accuracy of 0.1~mm, Case 3 is for Venus-Earth, Mars-Earth and Venus-Mars with an accuracy of 10$^{-3}$~mm}
\label{casAa}
\end{figure}

\section{Configuration B: 2.5 years} 

By considering half the duration considered on Fig. \label{casAa}, we reduce significantly the performances. 
For Case 1 and 2 (without and with Venus-Mars distances at an accuracy of 0.1~mm respectively), we see that the introduction of the Venus-Mars measurements has a less important impact than over 5 years, especially for the main belt asteroid masses. The inner planet semi-major axis are improved two times less than over 5 years. For Case 3 (Venus-Earth, Mars-Earth and Venus-Mars with 10$^{-3}$~mm accuracy) the effect is not so important and impacts mainly the asteroid masses for which the improvement remain below the 3 order of magnitudes obtained with 5 years of observations. The same comment is true for the Sun oblateness.

Regarding $\frac{\dot{\mu}}{\mu}$,  as one can see on Table \ref{tab:log},  we  have a degradation of about a factor 3 in comparison to a 5 years mission with 10$^{-3}$~mm accuracy, 2 at 0.1~mm. The improvement brought by the addition of the Mars-venus measured distances is also less important (1.5) relative to the 5-year case.

\begin{figure}
\includegraphics[scale=0.5]{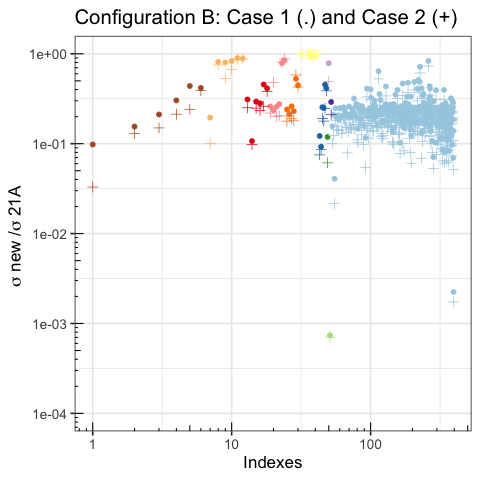}\includegraphics[scale=0.5]{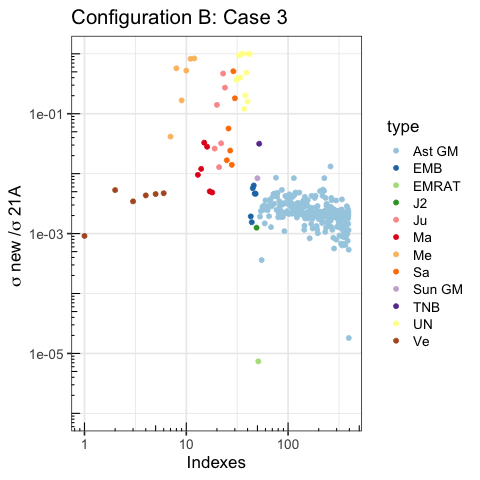}
\caption{Same as Fig. \ref{casAa} but in considering 2.5 years of observations  and a limit on the separation angle (SEP) between the Sun and the planets.}
\label{casB}
\end{figure}

\begin{table}
\centering
\caption{Ratio of the $\frac{\dot{\mu}}{\mu}$ uncertainties $\sigma_{new}/2\sigma_{21A}$ obtained for  different configurations and cases.}

\begin{tabular}{l c c c}
\hline
Configuration & case 1 & case 2 & case 3 \\
& VE-ME & VE-ME-VM & VE-ME-VM \\
& 0.1~mm  & 0.1~mm & 10$^{-3}$~mm \\
\hline
 A 5 yrs & 0.09  & 0.05 & 0.001  \\
 B 2.5 yrs & 0.15 & 0.10  & 0.003  \\
 C 1 yr & 0.26 & 0.20 & 0.009   \\
 \hline
\label{tab:log}
\end{tabular}
\end{table}

\section{Configuration C: 1 year} 

This configuration is the less effective as expected but still guarantees an improvement of about 2 orders of magnitude in average for the asteroid masses in the case 3 as well as for the sun oblateness and the orbits of the three observed inner planets.
For Case 1 and 2 (accuracy of 0.1~mm  without and with Venus-Mars distances respectively), very few parameters have an improvement close to one order of magnitude. In average, the improvement is of about a factor 5.  This comment is also true for $\frac{\dot{\mu}}{\mu}$  as one can see on Table \ref{tab:log}.
With a 10$^{-3}$~mm accuracy, it is still possible to have 2 orders of magnitude in gain for the asteroid masses, the Sun oblateness and the Venus, the EMB and some elements of the Mars orbits. The $\frac{\dot{\mu}}{\mu}$ improvement will be of about 2 orders of magnitude as well (see Table \ref{tab:log}).

\begin{figure}
\includegraphics[scale=0.5]{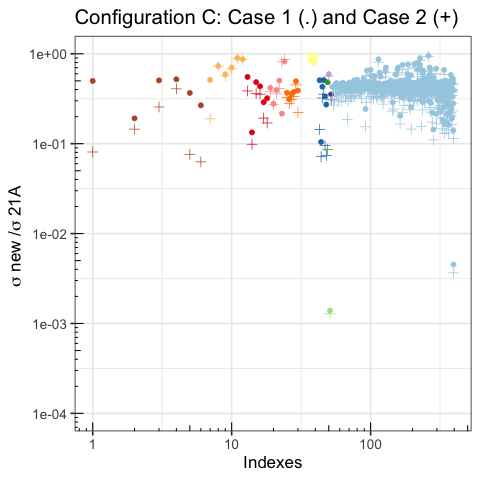}\includegraphics[scale=0.5]{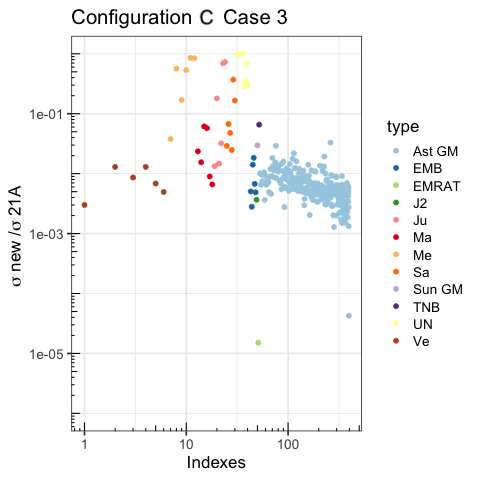}
\caption{Same as Fig. \ref{casAa} but in considering 1 year of observations and a limit on the separation angle (SEP) between the Sun and the planets .}
\label{casC}
\end{figure}

\section{Comparisons with Bepi-Colombo} 

\begin{figure}
 \centering
\includegraphics[scale=0.5]{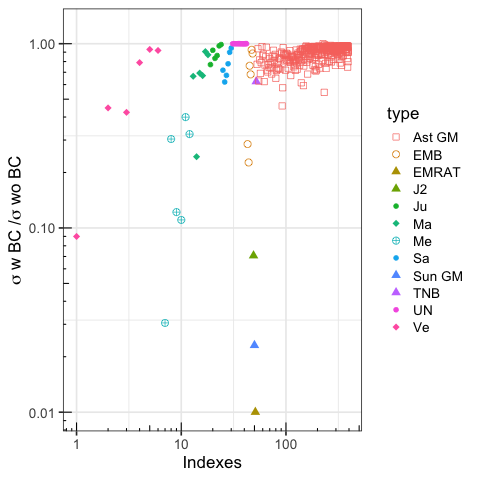}
\caption{Figure extracted from \cite{fienga2022}: Distribution of the ratio between the parameter uncertainties obtained with and without BC simulations in log scale.The colours and shapes indicate the different types of parameters considered in the INPOP adjustment : Me,V,Ma,Ju,Sa,UN, EMB represent the ratio of the uncertainties for the 6 orbital initial conditions for Mercury, Venus, Mars, Jupiter, Saturn, Uranus, Neptune and the Earth-Moon barycenter respectively. J2 , Sun GM, EMRAT and TNB give the ratios for the Sun oblateness and mass, the ratio between the Earth and the Moon masses and the mass of the TNO ring respectively. Finally Ast GM indicate the ratio for the 343 Main Belt asteroid masses.}
\label{fig:BC}
\end{figure}

\cite{fienga2022} gave similar comparisons to INPOP21a accuracy induced by the Bepi-Colombo (BC) MORE experiment. This experiment provides measurements of the Bepi-Colombo Earth distances at the 1~cm level thanks to a new generation of radio transponder in  Ka-Ka band. For these simulations a duration of 2 years has been considered. These measurements will benefit to the Mercury orbit computation as well as the full planetary ephemeris as explained in \cite{fienga2022}.
In order to compare the TLN results will the one simulated with BC we reproduce the figure from \cite{fienga2022} on Fig. \ref{fig:BC}.
In comparing Fig. \ref{casC} and \ref{fig:BC}, one can see that a 1-year mission with 0.1~mm accuracy,  representing two order of magnitude improvement relative BC performances, will bring a averaged gain of a factor 5 on the asteroid masses and relative equivalent performances for Venus and EMB orbits. With Case 1 and 2 (with a 0.1~mm accuracy) , even for a 5-year TLNL mission, the Sun mass and oblateness appear to be more constrained is BC over 2 years of mission.
In the other hand, with a 10$^{-3}$~mm accuracy, all the parameters are derived with an accuracy better than the one reached with BC simulations, including the Sun mass and oblateness.

\section{Conclusion}
Interplanetary Laser Tri-lateration Network (ILTN)  seems to be a very promising option for the measurement of the distribution of mass in the Solar System and of $\frac{\dot{\mu}}{\mu}$. The addition of the interplanetary link between Venus and Mars is very efficient relative to the Venus-Earth and Mars-Earth combinations, and this what ever the mission duration or the accuracy of measurement.

\bibliographystyle{plain}
\bibliography{global}

\end{document}